\documentclass[aps,prd,reprint,superscriptaddress,nofootinbib]{revtex4-1}
\usepackage{graphicx}
\usepackage{tabularx}
\usepackage{epsfig}
\usepackage{amsmath,amssymb}
\usepackage{verbatim}
\usepackage{lineno}
\usepackage{xspace}
\usepackage{dcolumn}
\usepackage{booktabs}  

\def\eq#1{\begin{linenomath}\begin{equation}#1\end{equation}\end{linenomath}}
\def \Nion {N_i^{\mathrm{on}}}
\def \Nioff {\langle N_i^{\mathrm{off}} \rangle}

\begin{document}

\title{Observation of the cosmic-ray shadow of the Moon with IceCube}

\affiliation{III. Physikalisches Institut, RWTH Aachen University, D-52056 Aachen, Germany}
\affiliation{School of Chemistry \& Physics, University of Adelaide, Adelaide SA, 5005 Australia}
\affiliation{Dept.~of Physics and Astronomy, University of Alaska Anchorage, 3211 Providence Dr., Anchorage, AK 99508, USA}
\affiliation{CTSPS, Clark-Atlanta University, Atlanta, GA 30314, USA}
\affiliation{School of Physics and Center for Relativistic Astrophysics, Georgia Institute of Technology, Atlanta, GA 30332, USA}
\affiliation{Dept.~of Physics, Southern University, Baton Rouge, LA 70813, USA}
\affiliation{Dept.~of Physics, University of California, Berkeley, CA 94720, USA}
\affiliation{Lawrence Berkeley National Laboratory, Berkeley, CA 94720, USA}
\affiliation{Institut f\"ur Physik, Humboldt-Universit\"at zu Berlin, D-12489 Berlin, Germany}
\affiliation{Fakult\"at f\"ur Physik \& Astronomie, Ruhr-Universit\"at Bochum, D-44780 Bochum, Germany}
\affiliation{Physikalisches Institut, Universit\"at Bonn, Nussallee 12, D-53115 Bonn, Germany}
\affiliation{Universit\'e Libre de Bruxelles, Science Faculty CP230, B-1050 Brussels, Belgium}
\affiliation{Vrije Universiteit Brussel, Dienst ELEM, B-1050 Brussels, Belgium}
\affiliation{Dept.~of Physics, Chiba University, Chiba 263-8522, Japan}
\affiliation{Dept.~of Physics and Astronomy, University of Canterbury, Private Bag 4800, Christchurch, New Zealand}
\affiliation{Dept.~of Physics, University of Maryland, College Park, MD 20742, USA}
\affiliation{Dept.~of Physics and Center for Cosmology and Astro-Particle Physics, Ohio State University, Columbus, OH 43210, USA}
\affiliation{Dept.~of Astronomy, Ohio State University, Columbus, OH 43210, USA}
\affiliation{Dept.~of Physics, TU Dortmund University, D-44221 Dortmund, Germany}
\affiliation{Dept.~of Physics, University of Alberta, Edmonton, Alberta, Canada T6G 2E1}
\affiliation{D\'epartement de physique nucl\'eaire et corpusculaire, Universit\'e de Gen\`eve, CH-1211 Gen\`eve, Switzerland}
\affiliation{Dept.~of Physics and Astronomy, University of Gent, B-9000 Gent, Belgium}
\affiliation{Dept.~of Physics and Astronomy, University of California, Irvine, CA 92697, USA}
\affiliation{Laboratory for High Energy Physics, \'Ecole Polytechnique F\'ed\'erale, CH-1015 Lausanne, Switzerland}
\affiliation{Dept.~of Physics and Astronomy, University of Kansas, Lawrence, KS 66045, USA}
\affiliation{Dept.~of Astronomy, University of Wisconsin, Madison, WI 53706, USA}
\affiliation{Dept.~of Physics and Wisconsin IceCube Particle Astrophysics Center, University of Wisconsin, Madison, WI 53706, USA}
\affiliation{Institute of Physics, University of Mainz, Staudinger Weg 7, D-55099 Mainz, Germany}
\affiliation{Universit\'e de Mons, 7000 Mons, Belgium}
\affiliation{T.U. Munich, D-85748 Garching, Germany}
\affiliation{Bartol Research Institute and Department of Physics and Astronomy, University of Delaware, Newark, DE 19716, USA}
\affiliation{Dept.~of Physics, University of Oxford, 1 Keble Road, Oxford OX1 3NP, UK}
\affiliation{Dept.~of Physics, University of Wisconsin, River Falls, WI 54022, USA}
\affiliation{Oskar Klein Centre and Dept.~of Physics, Stockholm University, SE-10691 Stockholm, Sweden}
\affiliation{Department of Physics and Astronomy, Stony Brook University, Stony Brook, NY 11794-3800, USA}
\affiliation{Dept.~of Physics and Astronomy, University of Alabama, Tuscaloosa, AL 35487, USA}
\affiliation{Dept.~of Astronomy and Astrophysics, Pennsylvania State University, University Park, PA 16802, USA}
\affiliation{Dept.~of Physics, Pennsylvania State University, University Park, PA 16802, USA}
\affiliation{Dept.~of Physics and Astronomy, Uppsala University, Box 516, S-75120 Uppsala, Sweden}
\affiliation{Dept.~of Physics, University of Wuppertal, D-42119 Wuppertal, Germany}
\affiliation{DESY, D-15735 Zeuthen, Germany}

\author{M.~G.~Aartsen}
\affiliation{School of Chemistry \& Physics, University of Adelaide, Adelaide SA, 5005 Australia}
\author{R.~Abbasi}
\affiliation{Dept.~of Physics and Wisconsin IceCube Particle Astrophysics Center, University of Wisconsin, Madison, WI 53706, USA}
\author{Y.~Abdou}
\affiliation{Dept.~of Physics and Astronomy, University of Gent, B-9000 Gent, Belgium}
\author{M.~Ackermann}
\affiliation{DESY, D-15735 Zeuthen, Germany}
\author{J.~Adams}
\affiliation{Dept.~of Physics and Astronomy, University of Canterbury, Private Bag 4800, Christchurch, New Zealand}
\author{J.~A.~Aguilar}
\affiliation{D\'epartement de physique nucl\'eaire et corpusculaire, Universit\'e de Gen\`eve, CH-1211 Gen\`eve, Switzerland}
\author{M.~Ahlers}
\affiliation{Dept.~of Physics and Wisconsin IceCube Particle Astrophysics Center, University of Wisconsin, Madison, WI 53706, USA}
\author{D.~Altmann}
\affiliation{Institut f\"ur Physik, Humboldt-Universit\"at zu Berlin, D-12489 Berlin, Germany}
\author{J.~Auffenberg}
\affiliation{Dept.~of Physics and Wisconsin IceCube Particle Astrophysics Center, University of Wisconsin, Madison, WI 53706, USA}
\author{X.~Bai}
\thanks{Physics Department, South Dakota School of Mines and Technology, Rapid City, SD 57701, USA}
\affiliation{Bartol Research Institute and Department of Physics and Astronomy, University of Delaware, Newark, DE 19716, USA}
% ...

\date{\today}

\collaboration{IceCube Collaboration}
\noaffiliation

\date{\today}

\keywords{Cosmic rays, Moon shadow, IceCube}
\begin{abstract}
 We report on the observation of a significant deficit of cosmic rays from the direction of the Moon
 with the IceCube detector. The study of this ``Moon shadow" is used to characterize the 
 angular resolution and absolute pointing capabilities of the detector. The detection is based on 
 data taken in two periods before the completion of the detector: between April 2008 and 
 May 2009, when IceCube operated in a partial configuration with 40 detector strings deployed in 
 the South Pole ice, and between May 2009 and May 2010 when the detector operated with 59 
 strings.
 Using two independent analysis methods, the Moon shadow has been observed to high significance 
 ($> 6\sigma$) in both detector configurations.  The observed location of the shadow center is within 
 $0.2^{\circ}$ of its expected position when geomagnetic deflection effects are taken into account. This
 measurement validates the directional reconstruction capabilities of IceCube.
\end{abstract}

\pacs{96.50.S-,95.85.Ry,96.20.-n,91.25.-r,29.40.Ka}

\maketitle

\section{Introduction}\label{sec:intro}

IceCube is a km$^{3}$-scale Cherenkov detector deployed in the glacial ice at the geographic South Pole. Its primary
goal is to search for astrophysical sources of high-energy neutrinos. A major background for this search is the high rate of 
atmospheric muons produced when cosmic rays with energies above a few TeV interact with the Earth's atmosphere.  The rate of muon events
in IceCube above several hundred GeV dominates the total trigger rate of the detector, and is approximately six orders
of magnitude higher than the rate of neutrino-induced events.

The incoming direction of multi-TeV cosmic muons is on average within 0.1$^\circ$ of arrival direction of the
primary cosmic-ray particle~\cite{pTmuons2013}. This implies that the distribution of incoming muons should mimic the almost isotropic
distribution of TeV cosmic rays in the sky \cite{LargeScaleAnis, SmallScaleAnis}. An important feature of the angular 
distribution of cosmic rays is the
presence of a relative deficit in the flux of cosmic rays coming from the direction of the Moon.  This effect, due to
the absorption of  cosmic rays by the Moon, was first predicted by Clark in 1957 \cite{Clark1957}, and its observation
has been used by several experiments as a way of calibrating the angular resolution and the pointing
accuracy of their particle detectors (see~\cite{Ambrosio2003b}, \cite{Achard2005}, \cite{Oshima2010},
\cite{Adamson2011}, or \cite{Bartoli2011} for recent results.)

For IceCube, the Moon shadow analysis is a vital and unique verification tool for the track reconstruction algorithms that are
used in the search for point-like sources of astrophysical neutrinos \cite{Abbasi2011a}, among other analyses. In this
paper we will report on the observation of the Moon shadow using data taken between April 2008 and May 2010, before the
completion of the IceCube Neutrino Observatory in December 2010.

Two independent analysis methods were used in the search for the Moon shadow. The first analysis performs a binned,
one-dimensional search for the Moon shadow that compares the number of events detected from the direction of the Moon to
the number of background events recorded at the same declination as the Moon but at a different right ascension. The
second method uses an unbinned, two-dimensional maximum likelihood algorithm that retrieves the best fit value for the
total number of events shadowed by the Moon.

Both methods show consistent results, and constitute the first statistically significant detection of the shadow of the
Moon using a high-energy neutrino telescope.

\section{Detector Configuration and Data sample}\label{sec:det_data}

\subsection{The IceCube detector}\label{subsec:det}

The IceCube neutrino telescope uses the deep Antarctic ice as a detection medium. High-energy
neutrinos that interact with nucleons in the ice produce relativistic leptons that emit Cherenkov
radiation as they propagate through the detector volume.  This Cherenkov light is detected by a
volumetric array of 5160 Digital Optical Modules (DOMs) deployed at depths between 1450 m and 2450 m
below the ice surface. Each DOM consists of a 25 cm diameter photomultiplier tube
(PMT)~\cite{Abbasi:2010vc} and the electronics for signal digitization~\cite{Abbasi:2008ym} housed
inside a pressure-resistant glass sphere.

The DOMs are attached to 86 strings that provide mechanical support, electrical power, and a data
connection to the surface.  Consecutive DOMs in each string are vertically separated by a distance
of about 17 m, while the horizontal spacing between strings is about 125 m.  A compact group of eight
strings with a smaller spacing between DOMs is located at the bottom of the detector and forms
DeepCore~\cite{DeepCorePaper}, which is designed to extend the energy reach of IceCube to lower neutrino energies. The
IceTop surface array, devoted to the detection of extensive air showers from cosmic rays with
energies between 300 TeV and 1 EeV, completes the instrumentation of the observatory.

\begin{figure}[t]
  \begin{center}	
    \includegraphics[width=0.35\textwidth]{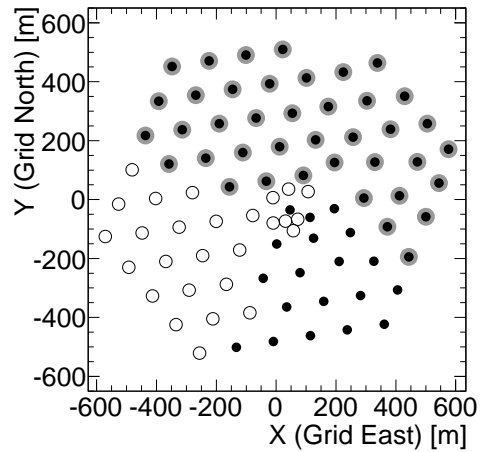}
    \caption{Layout of the two detector configurations considered in this analysis. IC40
    (\emph{gray}) operated between 2008 and 2009.  The deployment of more strings initiated the IC59
    configuration (\emph{black}) operated between 2009 and 2010. The remaining strings that
    form the final 86-string configuration, the last of them installed in December 2010, are shown
    as open circles. The $y$ axis (Grid North) is aligned with the Greenwich Prime Meridian.}
    \label{fig:ic40_ic59_layout}
  \end{center}
\end{figure}

The construction of IceCube began in 2005 and was completed in December 2010. During construction,
the detector operated in several partial configurations. Data from two different configurations were used
in this paper: between 2008 and 2009 the detector operated with 40 strings deployed in the ice
(IC40), and between 2009 and 2010 the detector operated in its 59-string configuration (IC59).
The layout of the two detector configurations used in this work can be seen in
Fig.~\ref{fig:ic40_ic59_layout}.
\subsection{Data sample}\label{subsec:data}

In order to reduce the rate of noise-induced events, IceCube DOMs are operated in a coincidence mode called
Hard Local Coincidence (HLC). During the operation of IC40 and IC59 the HLC requirement was met if photon hits were detected 
within a $\pm 1~\mu$s window in two nearest neighbor or next-to-nearest neighbor DOMs. The detection of HLC 
hits leads to a full readout and transmission to the surface of the digitized PMT signals. A trigger condition is then used
to combine these photon hits into a candidate event. 
The main trigger in IceCube is a simple multiplicity trigger called SMT8 that requires HLC hits in eight DOMs within 5 $\mu$s.
For each trigger, all HLC hits within a $\pm 10~\mu$s window are recorded and merged into a single event.

The majority of events detected by IceCube are due to down-going muons produced in the interaction
of high energy cosmic rays with the Earth's atmosphere. During the operation of IC40, the cosmic
muon-induced trigger rate was about 1.1 kHz, which increased to about 1.7 kHz during the IC59
data-taking period. This high rate of cosmic-ray muon events provides a high-statistics data set
that can be used to search for the Moon shadow. 

Since the rate of data transfer from the South Pole via the South Pole Archival and Data Exchange
(SPADE) satellite communication system is limited to about 100 Gb per day, only a limited number of 
muon events can be transmitted to North over the satellite.  For this reason, the data
used in this analysis were taken using a dedicated online filter that selects only events passing 
minimum quality cuts and reconstructed within a predefined angular acceptance window 
around the Moon.

A fast likelihood-based muon track reconstruction \cite{Ahrens:2003fg} is performed at the South
Pole to obtain the arrival direction of each event.  The reconstructed direction of the muon
track is then compared to the position of the Moon in the sky, which is calculated using the
publicly-available \texttt{SLALIB}\xspace library of astronomical routines \cite{slalib:2002}.

An event satisfies the Moon filter selection criterium if at least 12 DOMs in 3 different strings record
photon hits, and if the reconstructed direction is within 10$^{\circ}$ of the
Moon position in declination and 40$^{\circ} / \cos(\delta_{\mu})$  in right ascension (where
$\delta_{\mu}$ is the declination of the event and the cosine factor accounts for projection
effects.)  
 
\begin{figure}[t]
  \begin{center}	
    \includegraphics[width=0.45\textwidth]{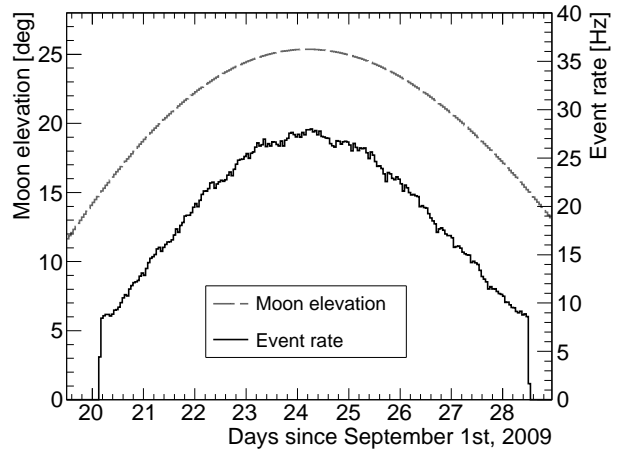}
    \caption{Rate of muon events passing the Moon filter during the month of September 2009, when
    IceCube was operating in its IC59 configuration. The correlation between Moon elevation
    (\emph{dashed line}) and event rate (\emph{solid line}) is clearly visible.} 
    \label{fig:rateplot}
  \end{center}
\end{figure}

The filter is enabled when the Moon is at least $15^{\circ}$ above the horizon. Due to the
particular geographic location of IceCube at the South Pole, the Moon rises above this threshold only
once per month, as its elevation above the horizon changes slowly over the course of days. Since the
number of muon events recorded by IceCube is a strong function of the elevation angle, the rate of
events that pass the acceptance window condition changes during this period as this window follows
the apparent motion of the Moon at the South Pole. The strong correlation between Moon elevation and
rate of events passing the Moon filter is shown in Fig.~\ref{fig:rateplot}. The maximum
event rate is also modulated over a longer time scale of 18.6 years (known as the lunar draconic
period \cite{1989EAA}) in which the maximum elevation of the Moon above the horizon at the South Pole oscillates
between the extreme values of $18.4^{\circ}$ and $28.4^{\circ}$. The maximum Moon elevation during
the IC40 data-taking period was $26.9^{\circ}$, while for IC59 was $25.6^{\circ}$.  Approximately
$1.29 \times 10^8$ muon events passing the Moon filter condition were recorded during the IC40
data-taking period, and about $1.77 \times 10^8$ events were recorded during the operation of the IC59
configuration. 

Once these events have been transferred from the South Pole, an iterative maximum-likelihood reconstruction
algorithm is applied to the data set to obtain a more precise track direction \cite{Ahrens:2003fg}.
The algorithm also determines the angular uncertainty in the reconstructed track direction by
mapping the likelihood space around the best track solution and fitting it with a paraboloid
function \cite{Neunhoffer2006220}. A narrow paraboloid indicates a precise reconstruction,
while a wide paraboloid indicates a larger uncertainty in the reconstructed direction of the muon
track. The $1\sigma$  contour line of the paraboloid function defines an error ellipse for the
reconstructed direction of the track. In this analysis, a single, one dimensional estimator of the
uncertainty is obtained by calculating the RMS value of the semi-major axes of that error
ellipse.

The likelihood-based track reconstruction algorithm used in this work is based on the
leading-edge times of the first light pulses recorded by each DOM.
For the fast track reconstruction at the South Pole the single-photo-electron (SPE) fit is used.  In
this fit, the likelihood that the first photon arrived at the pulse leading-edge times is maximized.  The photons
arriving at later times are ignored.

Neutrino point source searches rely on the multi-photo-electron (MPE) fit.  In the MPE fit, the total
number of photo-electrons $N_d$ in each DOM $d$ is taken into account by multiplying the likelihood
that a photon was detected at the first leading edge time with the probability that the remaining
$N_d-1$ photons arrived later \cite{Ahrens:2003fg}.  For bright, i.e. high-energy, events in simulated data the MPE fit
results in a slightly better angular resolution than the SPE fit.  Also the number of direct
(unscattered) photons associated with the reconstructed track tends to be larger with the MPE fit
than with the SPE fit.  This makes this quantity as well as related quantities more effective for
selecting well-reconstructed events. The MPE fit is discussed further in Section~\ref{subsubsec:mperesults}.

The track reconstruction algorithms use the local detector coordinate system and the direction of a
reconstructed track is given as a zenith and azimuth angle.  Using the event times as recorded by the
data acquisition system these are transformed into a right ascension $\alpha_\mu$ and declination
$\delta_\mu$, which are the more natural variables for searches of neutrino point-like
sources.

\section{Simulation}\label{sec:simulation}

\subsection{Cosmic ray energy and composition}\label{subsec:energy}

The muons produced in the interaction between the cosmic rays and the atmosphere must traverse
several kilometers of ice before reaching the IceCube detector, losing energy in the process. This
sets a lower limit of several hundred GeV on the energy of the muons at ground level that would
trigger the detector. By extension, the primary cosmic-ray particle needed to produce this kind of
muon should have an energy of at least several TeV. In the following, we will refer to the energy of
the primary cosmic ray, not the muons, unless specified otherwise. 

Given that this analysis deals with cosmic-ray showers near the energy threshold of the detector, the number of muons 
produced in each shower that reach the detector is small. Most events in the Moon data sample are composed of one or 
two energetic muons, and only 2\% of the events have muon multiplicities higher than ten.

The detailed energy scale for the IC40 and IC59 data sets was determined using simulated cosmic ray
air showers created with the CORSIKA Monte Carlo code~\cite{Heck:1998a} using the SIBYLL model of
high-energy hadronic interactions \cite{Ahn:2009wx}. The chemical composition and spectral shape of
the cosmic rays generated in this simulation follow the polygonato model \cite{Hoerandel:2002yg}.
 
From these simulations, we estimate that the median energy of the primary cosmic rays that trigger the
IceCube detector is 20 TeV, while the median energy of events that satisfy the Moon
filter condition is about 40 TeV for both IC40 and IC59, with 68\% of the events between
10 TeV and 200 TeV.  The increased median energy of the filtered sample is due to the greater average
zenith angles of the cosmic rays that pass the filter, which requires primary particles with enough
energy to produce muons able to traverse more ice and trigger IceCube. The muons produced by cosmic 
rays passing the Moon filter have a mean energy of about 2 TeV at ground level and reach the detector 
with a mean energy of 200 GeV. 
\begin{figure}[t]
  \begin{center}	
    \includegraphics[width=0.45\textwidth]{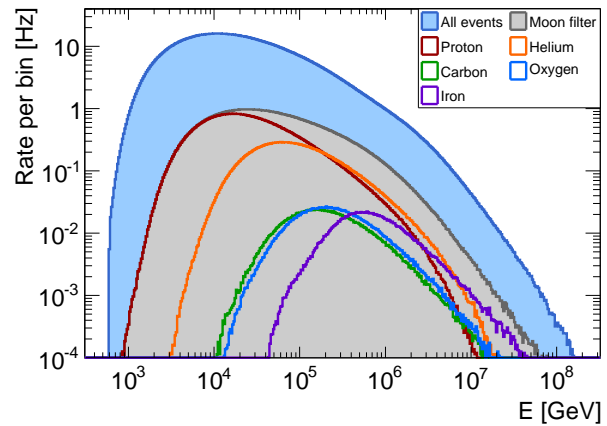}
    \caption{Differential event rate as a function of cosmic-ray primary energy for all events in IC59 (\emph{light
    blue}) and for only those passing the Moon filter (\emph{gray}) as determined from simulation studies. The main chemical elements that
    make up the events passing the Moon filter are shown with lines of different color. The width of the energy
    bins is 0.014 in $\log_{10}(E)$.The IC40 configuration shows a similar energy response.}
    \label{fig:ic59_spectrum}
  \end{center}
\end{figure}

The energy spectrum of all primary cosmic rays triggering the IceCube detector is shown in
Fig.~\ref{fig:ic59_spectrum} and compared to the spectrum of those that pass the Moon filter.  Also
shown in the figure are the five main chemical elements (protons, He, C, O, and Fe) that make more
than 95\% of the Moon filter data sample assuming the polygonato composition model.
The two main components of the sample are proton (68\% of the events) and helium (23\%).
 
\subsection{Geomagnetic field effects}\label{subsec:geomag}

Cosmic rays with TeV energies should experience a small deflection in their
trajectories due to the influence of the magnetic field of the Earth as they propagate towards the
detector. This deflection would appear in the Moon shadow analysis as a shift in the position of
the shadow with respect to the true Moon position, which could be wrongly interpreted as a systematic offset produced by the
event reconstruction. 

In order to quantify this offset and compare it with any possible shift observed in the data, we
have developed a particle propagation code that can be used to trace cosmic rays in the geomagnetic
field. Using this code, particles are propagated radially outwards from the South Pole up to a
a distance of 30 Earth radii from the center of the Earth at which point the opening angle between the initial and final
velocity vectors is computed. This angle gives the magnitude of the deflection in the geomagnetic
field.

We use the International Geomagnetic Reference Field (IGRF) model \cite{GJI:GJI4804} to
calculate deflections. In this model, the field is calculated using a truncated multipole series expansion. The current revision of the
model, IGRF-11, can be used to calculate B-field values through 2015, providing a good coverage of
the time range over which the data were taken. The model is accessible through a library of
\texttt{FORTRAN} routines called \texttt{GEOPACK} developed by
N.~Tsyganenko\footnote{\href{http://geo.phys.spbu.ru/~tsyganenko/modeling.html}
{\path{http://geo.phys.spbu.ru/~tsyganenko/modeling.html}}}.

The IGRF-11 model describes what is known as the \emph{internal} magnetic field of the Earth, which
is presumably produced by electric currents in the outer core of the planet and accounts for most of
the total magnetic field. A weaker component, known as the \emph{external} field, is produced by
electrical currents in the ionosphere. The external component is not included in our
calculation since it only modifies the total angular deflection by a few percent while significantly
increasing the computation time needed to perform the simulation.

\begin{figure}[t]
  \begin{center}	
    \includegraphics[width=0.45\textwidth]{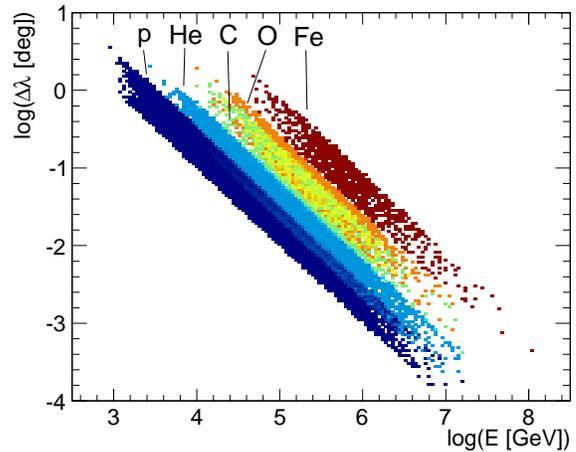}
    \caption{Angular deflection as a function of energy for the different chemical elements
    simulated using the particle propagation code described in Section~\ref{subsec:geomag}.}
    \label{fig:comp}
  \end{center}
\end{figure}

In our simulation, primary cosmic rays are propagated in the direction of the Moon as seen from the
South Pole for different times during the data taking period. The cosmic ray energy and chemical
composition is sampled from the event distributions that pass the Moon filter, shown in
Fig.~\ref{fig:ic59_spectrum}.  The resulting total deflection $\Delta\lambda$ is shown in
Fig.~\ref{fig:comp} as a function of energy for $10^{5}$ simulated cosmic ray particles for the five
main chemical elements that contribute to the Moon dataset. The energy and charge dependence of the
deflection angle is evident in the plot. Different bands in the plot correspond to different chemical elements. The width of each band is due 
to particles that were propagated in 
different directions in the sky (i.e. through different regions of the Earth's magnetic field) experiencing different
deflections. 
A power-law fit to the simulation results has been performed to estimate the deflection angle as a 
function of energy and charge. The fit gives a good agreement for the following expression:
\eq{ \Delta\lambda \mathrm{[^{\circ}]}= 1.9^{\circ} \frac{Z}{E \mathrm{[TeV]}} ~~, }
where $Z$ is the CR charge in units of elementary charge $e$, $E$ is its energy in TeV, and
$\Delta\lambda$ is given in degrees. This expression has the same functional form as the one found
in \cite{DiSciascio2011301} with a higher normalization in our simulation, which could be
due to the difference in geographic location and other simulation details.

The deflection of each cosmic ray with arrival direction $(\alpha_{\mu}, \delta_{\mu})$ in sidereal coordinates 
is calculated with respect to the position of the Moon at the time of the event $(\alpha_{\mathrm{Moon}}, \alpha_{\mathrm{Moon}})$. The two coordinates that characterize the position of an event in this
system are a right ascension difference $\Delta\alpha = (\alpha_\mu - \alpha_\mathrm{Moon}) \cos\delta_\mathrm{\mu}$, and a declination difference  $\Delta\delta = \delta_\mu - \delta_\mathrm{Moon}$ with respect to the nominal Moon position. 
The median shift in right ascension $\Delta\alpha$ for all CR particles in our simulation is $0.08^{\circ}$, 
with 68\% of the particles having deflection angles in the interval $0.02^{\circ} < \Delta\alpha < 0.24^{\circ}$. 
The median shift in declination $\Delta\delta$ is consistent with $0^{\circ}$,
with 68\% of the events contained in the interval $| \Delta\delta | < 0.035^{\circ}$. 

The cosmic-ray muons that ultimately trigger IceCube are also deflected by the geomagnetic
field. However, since their total track length is in the 50-100 km range and their energy is about 2 TeV, their contribution
to the total deflection angle should be at most $\sim 0.015^{\circ}$. For this reason, the muon contribution has been
ignored in calculating the expected total deflection angle.

\section {Binned analysis}\label{subsec:ana_b}

\subsection{Description of the method}

The main goal of the binned analysis is to obtain a profile view of the Moon shadow and measure its width, 
which can be used as a direct estimator of the angular resolution of the event reconstruction. This is
accomplished by comparing the observed number of events as a function of angular distance from the Moon
to an estimate of how many events would have been observed if there was no shadow. 

For this comparison, the angular distance between the reconstructed muon tracks and the expected position of 
the Moon is binned in constant increments of $0.2^{\circ}$ up to a maximum angular distance of $5^{\circ}$. 
This defines the so-called \emph{on-source} distribution of events. The same binning procedure is applied to 
eight \emph{off-source} regions centered around points located at the same declination as the Moon, but offset 
from it in right ascension by $\pm 5^{\circ}$, $\pm10^{\circ}$, $\pm15^{\circ}$, and $\pm 20^{\circ}$, where it is 
assumed that the shadowing effect is negligible. The average number of counts as a function of radius for these 
eight \emph{off-source} regions represents the expectation in the case of no Moon shadow.

The relative difference between the number of events in the $i$-th bin in the \emph{on-source} region $\Nion$, 
and the average number of events in the same bin in the \emph{off-source} regions $\Nioff$ 
is calculated using the following expression:
\eq{
\frac{\Delta N_i}{\langle N \rangle_i} = \frac{\Nion - \Nioff}{\Nioff} ~~.
}
The uncertainty in the relative difference is given by:
\eq{
	\sigma_{\Delta N/{\langle N \rangle}} = \frac{\Nion}{\Nioff} \sqrt{ \frac{1}{\Nion} + \frac{1}{s \Nioff} } ~~,
} 
where $s = 8$ is the number of \emph{off-source} regions. The distribution of relative differences as a function 
of angular radius from the Moon constitutes a profile view of the shadow. 

Simulation studies indicate that the point spread function (PSF) of the detector can be approximated with 
a two-dimensional Gaussian function.  We use this approximation to obtain an estimate of the angular resolution
of the track reconstruction by fitting the distribution of $\Delta N_i / \langle N \rangle_i$ for the events in the Moon 
data set.

Following \cite{Cobb2000}, we treat the Moon as a point-like cosmic ray sink that removes $\Phi \pi R_M^2$ events
from the muon sample, where $R_M$ is the angular radius of the Moon ($R_M \sim 0.26^{\circ}$) and $\Phi$ is the 
cosmic ray flux at the location of the Moon in units of events per square degree. This deficit is smeared by the PSF of 
our detector, which results in radially-symmetric two-dimensional Gaussian distribution of shadowed events which is
 a function of radial distance $\psi$ from 
the center of the Moon. Integrating over the azimuthal coordinate of the symmetric Gaussian distribution yields:
\eq{
	f(\psi) = - \frac{\Phi \pi R_M^2}{\sigma^2} e^{-\psi^2 / 2 \sigma^2} ~~,
}
where $\sigma$ is the angular resolution of the directional reconstruction. The number of shadowed events in the 
$i$-th bin of width $\Delta \psi$ is given by the two-dimensional integral:
\begin{eqnarray}
N_s(\psi_i) &=& \int^{\psi_i + \Delta \psi/2}_{\psi_i - \Delta \psi/2} \psi \; f(\psi) \; d\psi \\
&\approx& - \frac{\Phi \pi R_M^2 \Delta \psi}{\sigma^2} \psi_i \; e^{-\psi_i^2 / 2 \sigma^2} ~~. \label{ns}
\end{eqnarray}

The number of events $N_e$ that would have been observed in the same bin with no shadowing is $2 \pi \Phi \psi_i \Delta \psi  \label{ne}$.

The ratio of equations \ref{ns} and $N_e$ gives us the expected distribution of relative differences $\Delta N_i /
\langle N \rangle_i$ for a detector with a Gaussian PSF of angular resolution $\sigma$: 
\eq{
	\frac{N_s}{N_e}(\psi_i) = - \frac{R_M^2}{2 \sigma^2} e^{-\psi_i^2 / 2 \sigma^2} ~~. \label{gaussfit}
}
This expression is used to fit the experimental data, and the resulting value of $\sigma$ is compared to an estimate
of the angular resolution of the data set obtained from simulation studies. Our treatment ignores the finite angular size of the lunar disc, which may 
affect the result of the fit. However, for the expected angular resolution of the Moon data set (of order $1^{\circ}$ or less) this 
effect should influence the fit value of $\sigma$ only at the few-percent level. 

A set of cuts was developed to optimize for the statistical significance of the detection of the
Moon shadow. Under the assumption of Poisson statistics, the relation between the significance $S$,
the fraction $\eta$ of events passing the cuts, and the resulting median angular resolution
$\Psi_\mathrm{med}$ after cuts is: 

\eq{
 S \propto \frac{\sqrt{\eta}}{\Psi_\mathrm{med}} ~~.
}

The optimization of the cuts was performed on the CORSIKA-simulated air showers described in Section
\ref{sec:simulation}. Two cut variables were used in this analysis: the angular uncertainty
$\sigma_i$ in the reconstruction of the muon track direction estimated individually for each event, and the reduced
log-likelihood $rlogl$, which is the log-likelihood for the best track solution divided by the
number of degrees of freedom in the fit. The number of degrees of freedom in the track fit is equal
to the number of DOMs triggered by the event minus the number of free parameters in the fit (five
for this fit.) Both $rlogl$ and $\sigma$ are standard cut variables used in the search for point-like
 sources of astrophysical neutrinos \cite{Abbasi2011a}, the search for a diffuse flux of high-energy 
 neutrinos \cite{PhysRevD.84.082001}, and several other analyses of IceCube data.

Once the cuts have been determined, the number of events falling inside a circular search bin around
the Moon is compared to the number of events contained in a bin of the same angular radius for the
average \emph{off-source} region. The statistical significance of an observed deficit in the number of events 
in the search bin is calculated using the method given by \cite{LiMa1983}.

The optimal radius of the search bin $\psi_b$ can be found by maximizing the $S'$ parameter in the following 
expression:
\eq{
 S'(\psi_s) \propto \frac{\int^{\psi_s}_0 \psi'~\mathrm{PSF}(\psi')~d\psi'}{\psi_s} ~~,
}
where $\psi_s$ is the radius of the bin and $\mathrm{PSF}(\psi')$ is the point spread function of the detector after cuts
obtained from simulations. Due to its symmetry, the PSF has already been integrated over the azimuthal coordinate and 
only the radial dependence remains. The optimization of the search bin radius is also performed using simulated CORSIKA 
showers generated for each detector configuration. 

\subsection{Results\label{subsec:results}}

A set of cuts was determined independently for both the IC40 and IC59 detector configurations using the optimization
procedure described above on simulated data. For IC40, only events with $rlog < 9$ and $\sigma_i <
1.01^{\circ}$ were used in the analysis, with 26\% of the events surviving the cuts. After cuts, the median 
angular resolution of the reconstruction was estimated from simulation to be $0.93^{\circ}$, with 68\% of the 
events having angular uncertainties $\sigma_i$ between $0.38^{\circ}$ and $2.18^{\circ}$. A two-dimensional fit to the 
simulated data shows that for the Gaussian approximation the corresponding resolution $\sigma$ is about
$0.74^{\circ}$.

In the case of IC59, the events selected for the analysis were those with $rlog < 8.8$ and $\sigma_i
<1.04^{\circ}$, which resulted in a passing rate of 34\%. The median resolution after cuts was
$0.78^{\circ}$, with the 68\% containing interval located between $0.33^{\circ}$ and
$1.78^{\circ}$, with a Gaussian width $\sigma$ of about $0.71^{\circ}$. 

After the cuts were applied to both data sets, the radius of the optimal search bin ($\psi_b$) and
the number of events contained in that bin for both the \emph{on-source} ($N_{\mathrm{on}}^b$), and
\emph{off-source} ($N_{\mathrm{off}}^b$) windows were calculated. In both detector configurations, a
deficit in the number of events in the \emph{on-source} bin when compared to the \emph{off-source}
bin was observed at high statistical significance ($> 6\sigma$), as expected due to the shadowing
effect of the Moon. A complete list of the number of events observed on each bin, the observed
deficit in the \emph{on-source} bin, as well as the statistical significance associated with such
deficit is given in Table~\ref{tab}.

\begin{table}[htp]
\centering
\begin{tabular}{p{4cm}   p{2cm}   p{2cm}}
\hline
\hline
 & \bf{IC40} & \bf{IC59} \\
\hline
$\psi_b$ & $0.75^{\circ}$ & $0.79^{\circ}$  \\
$N_{\mathrm{on}}^b$ & 52967 & 96412 \\
$N_{\mathrm{off}}^b$ & 54672 & 100442 \\
$\Delta N$ & -1705 & -4030 \\
Significance & $6.9\sigma$ & $12.1\sigma$ \\
\hline
\end{tabular}
\caption{\label{tab} 
 Optimal bin radius ($\psi_b$), number of observed events in the \emph{on-source}
 ($N_{\mathrm{on}}^b$) and \emph{off-source} ($N_{\mathrm{off}}^b$) bins,  Event deficit in the
 \emph{on-source} bin ($\Delta N$), and statistical significance of the deficit for the binned
 analysis of IC40 and IC59 data sets.
}
\end{table}

\begin{figure*}[t]
  \begin{center}	
  $\begin{array}{cc} %
  \includegraphics[width=.4\textwidth]{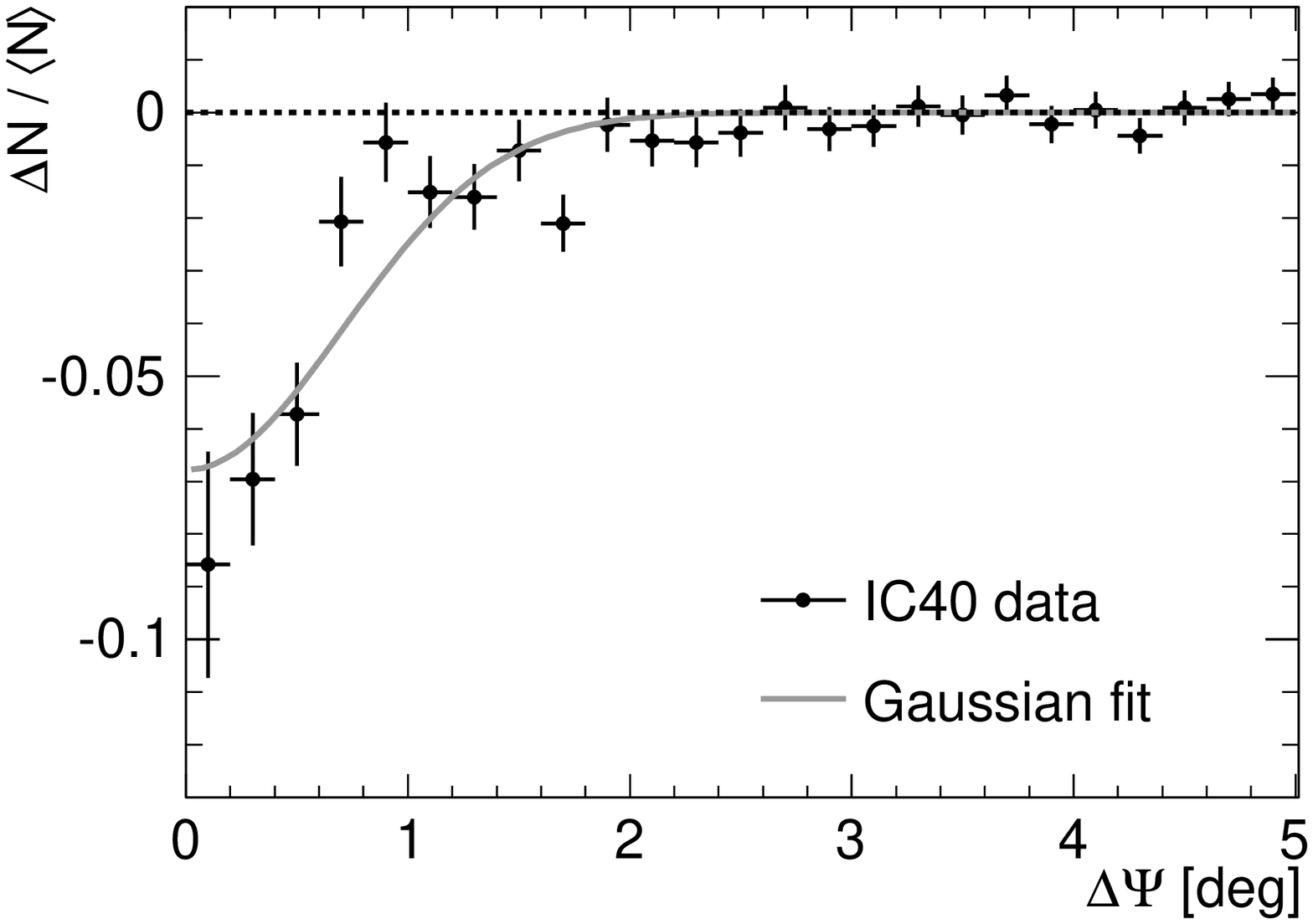} & %
   \includegraphics[width=.4\textwidth]{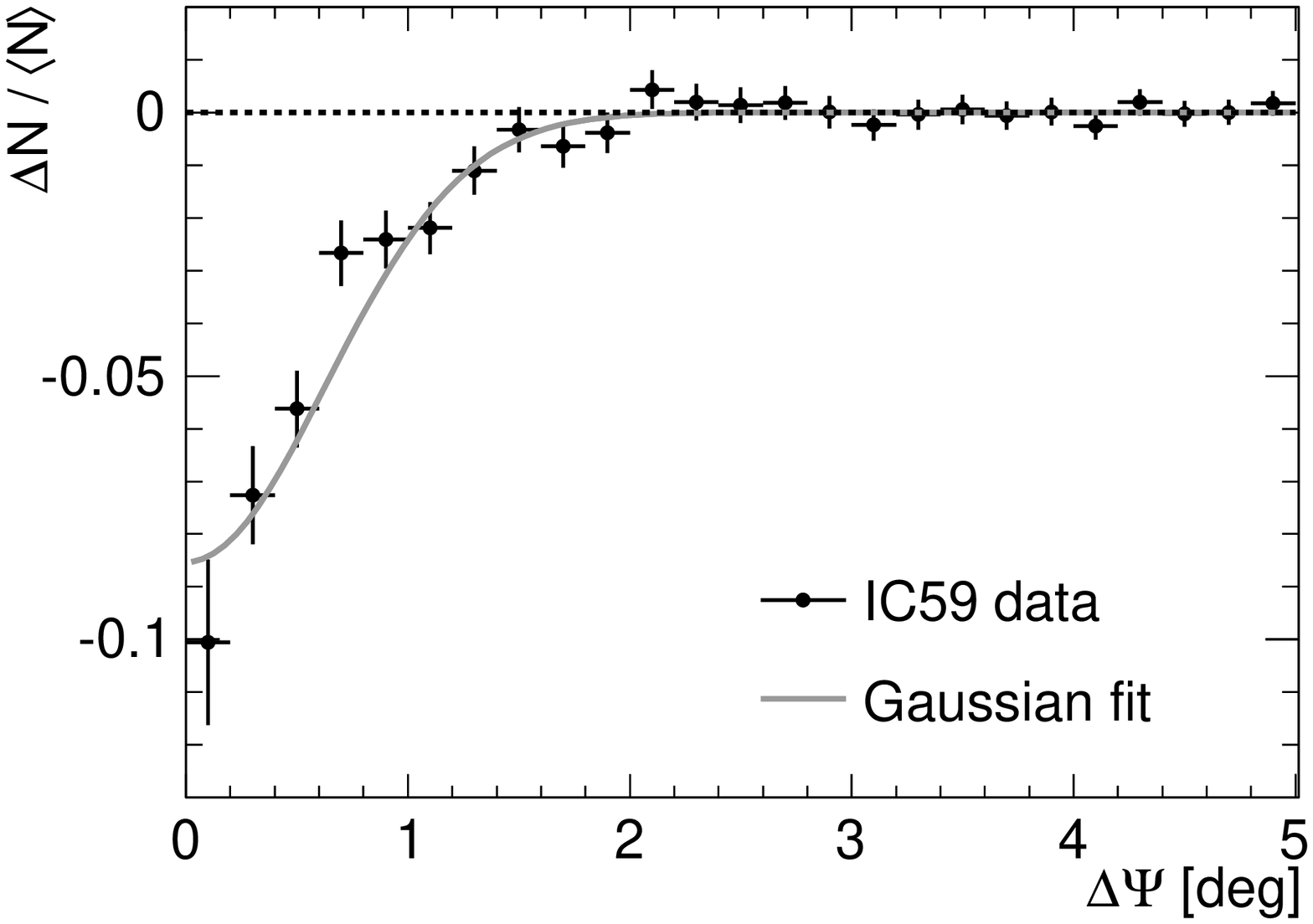} %
   \end{array}$ %
   \end{center} %
    \caption{ Relative difference between the number of events in the \emph{on-source} and the
    average \emph{off-source} region as a function of the angular distance from the nominal position
    of the Moon for the IC40 \emph{(left)} and IC59 \emph{(right)} datasets. A Gaussian fit to the
    deficits is shown in gray.} 
    \label{fig:binned}
\end{figure*}

The Moon shadow profile shown in Fig.~\ref{fig:binned} was fit using the expression
given in Eq.~\ref{gaussfit}, where $\sigma$ is the only free parameter. A list of fit
results is given in Table~\ref{tab_b}. In both cases, the observed angular resolution shows good agreement with the
one obtained from the above-mentioned simulation studies. 

\begin{table}[htp]
\centering
\begin{tabular}{p{2cm}   p{3cm}   p{3cm}}
\hline
\hline
 & \bf{IC40} & \bf{IC59} \\
\hline
$\sigma$ & $0.71^{\circ} \pm 0.07^{\circ}$ & $0.63^{\circ} \pm 0.04^{\circ}$ \\
$\chi^2 / \mathrm{dof}$ & 31.4 / 24 & 13.0 / 24 \\
\hline
\end{tabular}
\caption{\label{tab_b} 
 Gaussian angular resolution $\sigma$ obtained from the fit to the Moon shadow profile shown in
 Fig.~\ref{fig:binned}. The $\chi^2 / \mathrm{dof}$ of the fit is also given for the two results.
}
\end{table}
%

\section {Unbinned analysis}\label{subsec:ana_ub}

\subsection{Description of the method}

The second algorithm used to search for the Moon shadow is based on an
unbinned maximum likelihood method analogous to that used in the search for point-like sources of
high-energy neutrinos~\cite{Braun2008299}. This kind of likelihood analysis was first proposed in
\cite{PhysRevD.25.2846}, and was applied for the first time to a Moon shadow search in
\cite{Wascko00}.

The goal of the unbinned analysis is to determine the most likely location of the Moon shadow
to compare it with the expected location after accounting for magnetic deflection effects. An agreement
between the observed and expected positions of the shadow center will serve as an important
confirmation of the absolute pointing accuracy of the detector. 

The analysis is also used to obtain the most likely number of events shadowed by the Moon, 
which can be compared to the expectation. An essential ingredient in the unbinned analysis
is an event-wise estimation of the angular error. Both systematic under- and overestimation of
this error would lead to a shallower apparent shadow than expected. The number of shadowed
events is a free parameter in this analysis and the comparison with the expected number of 
shadowed events is effectively a test of the angular uncertainty estimate.

In this analysis~\cite{Blumenthal:2011,Reimann:2011}, the position of each muon event is defined 
with respect to the Moon position in the coordinate system $(\Delta\alpha, \Delta\delta)$ that was
defined in Section~\ref{subsec:geomag}. Only events with $|\Delta\delta|
\le 8^{\circ}$ and $|\Delta\alpha + \alpha_\mathrm{off}| \le 8^{\circ}$ were considered in the
analysis, where $\alpha_\mathrm{off} = 0^{\circ}$ defines the \emph{on-source} region, and
$\alpha_\mathrm{off} = \pm 18^{\circ}$ define two \emph{off-source} regions. 

A set of quality cuts was determined for this analysis using the same simulation dataset as in the
one-dimensional binned case. The same variables were used in the optimization of the cuts: the
angular reconstruction uncertainty $\sigma_i$, and the reduced log-likelihood of each event
\emph{rlogl}. 

The analysis method assumes that the data can be described as a linear combination of signal and
background components, where the relative contribution from each component is established by a
maximum likelihood fit to the data.  For a data set containing $N$ events, the log-likelihood
function is defined as
 \eq{
\log {\cal L}(n_s,\vec{x}_s) = \sum^N_{i=1} \log \Big[ \frac{n_s}{N} {\cal S}(\vec{x}_i,\sigma_i;\vec{x}_s) + (1 - \frac{n_s}{N}) {\cal B}(\vec{x}_i) \Big] ~~,
\label{eq:llhmoon}
 } 
where ${\cal S}$ and ${\cal B}$ are the signal and background probability density functions (PDFs),
$n_s$ is the unknown number of signal events, or in this case the total number of shadowed events,
and $\vec{x}_s$ is the unknown central position of the shadow of the Moon, relative to the nominal
position of the Moon.  Note that since the expected signal in the case of the Moon is a deficit in
the muon flux rather than an excess, $n_s$ should be negative. In absence of a geomagnetic
field, the shadow should occur exactly on the nominal position of the Moon, i.e.\ $\vec{x}_s=(0,0)$,
but according to the estimates described in section~\ref{subsec:geomag} we expect the shadow to be
shifted by about $0.1^\circ$.  

Other potential sources of systematic errors could also produce a shift or a smearing of the shadow.  
For example, it is expected that the timing accuracy of IceCube should be better than 1 $\mu$s, but if 
due to a detector malfunction a systematic error of a few minutes were introduced in the event registration time, the 
observed position of the shadow would experience a shift of order $0.1^\circ$. A similar effect 
would be observed if the detector geometry used for reconstructions differed from the real one by
several meters, or if the properties of light propagation in the ice \cite{SpicePaper} induced a systematic shift in the 
reconstructed direction of the muon tracks.
In fact, a measurement of the shadow of the Moon with the expected location, depth and width serves
as an independent verification that all these possible systematical errors are indeed under control.

The signal PDF for each event is modeled using a two-dimensional Gaussian distribution around the
reconstructed direction $\vec{x}_i$ of the muon track:
\eq{
 {\cal S}(\vec{x}_i,\sigma_i;\vec{x}_s) = \frac{1}{2 \pi \sigma^2_i} e^{-\frac{| \vec{x}_i - \vec{x}_s|^2}{2 \sigma_i^2}} ~~,
} 
where the width of the Gaussian distribution $\sigma_i$ is the angular reconstruction uncertainty
obtained on an event-by-event basis by the paraboloid algorithm described in
Section~\ref{subsec:data}.

The background PDF is assumed to depend only on $\Delta\delta$, and is derived from the distribution
of reconstructed declination angles for the muon tracks contained in the two \emph{off-source}
regions.

\begin{figure}[t]
  \begin{center}	
      \includegraphics[width=0.4\textwidth]{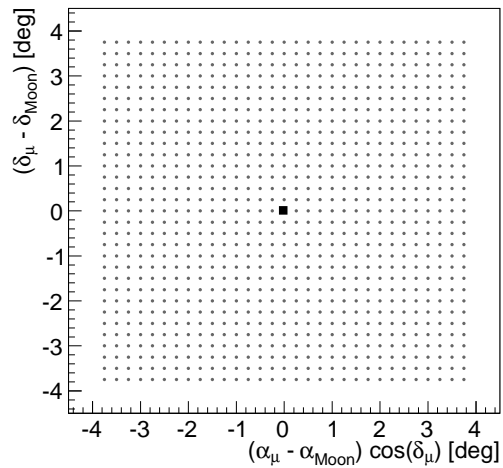} 
    \caption{Search grid in $(\Delta\alpha, \Delta\delta)$ as used in the unbinned likelihood
    analysis.  A value of $n_s$ is determined for each one of the points in the grid.  The nominal
    location of the Moon is shown as a black square at $\vec{x}_s$ = (0,0).}
    \label{fig:llhgrid} 
  \end{center}
\end{figure}

The best fit values for the number of signal events in the data $n_s$ and the shift of the shadow
center $\vec{x}_s$ are determined by maximizing the log-likelihood function (\ref{eq:llhmoon}).
Besides $\vec{x}_s$ and $n_s$ also the width and overall shape of the shadow are of interest. In
searches for point sources of high-energy neutrinos~\cite{Braun2008299}, for all points $\vec{x}_s$
on a fine grid covering the sky, the value of $n_s$
is determined which maximizes the likelihood function. Similarly, in the Moon shadow analysis we
determine the value of $n_s$ that maximizes the likelihood function (\ref{eq:llhmoon}) on a
rectangular grid of 961 values for $\vec{x}_s = (\Delta\alpha_s, \Delta\delta_s)$.  This
31$\times$31 grid is defined inside a window with a size of $|\Delta\delta| \le 4^{\circ}$ and
$|\Delta\alpha| \le 4^{\circ} $ shown in Fig.~\ref{fig:llhgrid}~.  

In order to avoid edge effects, all events in the $8^\circ\times8^\circ$ \emph{on-source} region are taken into account in the
maximum likelihood calculation.

The statistical significance associated with each value of $n_s$ can then be calculated by applying
the same likelihood analysis to the two \emph{off-source} regions. The RMS spread of the resulting
distribution of $n_s$ values for those regions gives an estimate of the $1\sigma$ spread expected in
the case of a null detection.  Using this estimate, each point in the \emph{on-source} region can be
given a statistical significance by taking the ratio between the value of $n_s$ at that point and
the $1\sigma$ estimate from the \emph{off-source} regions.

The observed value of $n_s$ is compared to an estimate of the true number of CRs
shadowed by the Moon. This estimate is obtained by counting the number of events that fall within a
circular window with the same radius as the Moon but located in the \emph{off-source} region.

\begin{table}[htp]
\centering
\begin{tabular}{p{4cm}   p{2cm}   p{2cm}}
\hline
\hline
 & \bf{IC40} & \bf{IC59} \\
\hline
Events before cuts & 18.8 $\times 10^6$ & 22.2 $\times 10^6$  \\
Cut 1: $ 0.075^{\circ} < \sigma_i < 1.5^{\circ}$ & $50 \%$ & $58 \%$ \\
Cut 2: $ 6.5 <$ rlogl $< 8$ & $89 \%$ &  $91 \%$ \\
Events after cuts & 8.4 $\times 10^6$ & 11.7 $\times 10^6$  \\
\hline
\end{tabular}
\caption{\label{tab1} Description of the cuts used in the unbinned analysis. Percentages indicate
the relative fraction of events that survive the cut with respect to the previous selection
criterium.}
\end{table}

\subsection{Results\label{subsec:ub_results}}

\subsubsection{SPE analysis}\label{subsubsec:speresults}

The cuts used in the unbinned analysis are listed in Table \ref{tab1}~. The resulting median
angular resolution of the IC40 and IC59 data sets was estimated by applying those same cuts to
simulated cosmic-ray events. In the case of IC40, the median angular resolution is $1.13^{\circ}$,
with 68\% of the events having angular uncertainties $\sigma_i$ between $0.48^{\circ}$ and $2.63^{\circ}$. For IC59, the median
resolution is $0.98^{\circ}$, with a 68\% containing interval defined between $0.38^{\circ}$ and
$2.23^{\circ}$.

\begin{table}[htp]
\centering
\begin{tabular}{p{4cm}   p{2cm}   p{2cm}}
\hline
\hline
 & \bf{IC40} & \bf{IC59} \\
\hline
Observed deficit  &  $5320 \pm 501$ & $8700 \pm 550$  \\
Expected deficit  &   $5734 \pm 76$   & $8192 \pm  91$  \\
Off-source RMS  &  521 &  627 \\
Significance       &  $10.2\sigma$ & $13.9\sigma$ \\
$\Delta \alpha$  & $-0.02^{\circ}$ &  $0.06^{\circ}$ \\
$\Delta \delta$  & $0.08^{\circ}$ &  $0.00^{\circ}$ \\
\hline
\end{tabular}
\caption{\label{tab_ub} Unbinned analysis results detailing the observed and expected deficit counts from 
the Moon for IC40 and IC59. The observed deficits and the ($\Delta\alpha, \Delta\delta$) offsets
are given for the most likely position of the Moon shadow as determined by the maximum likelihood fit.}
\end{table}

\begin{figure*}[t]
 \begin{center}
 $\begin{array}{cc} %
 \includegraphics[width=0.4\textwidth]{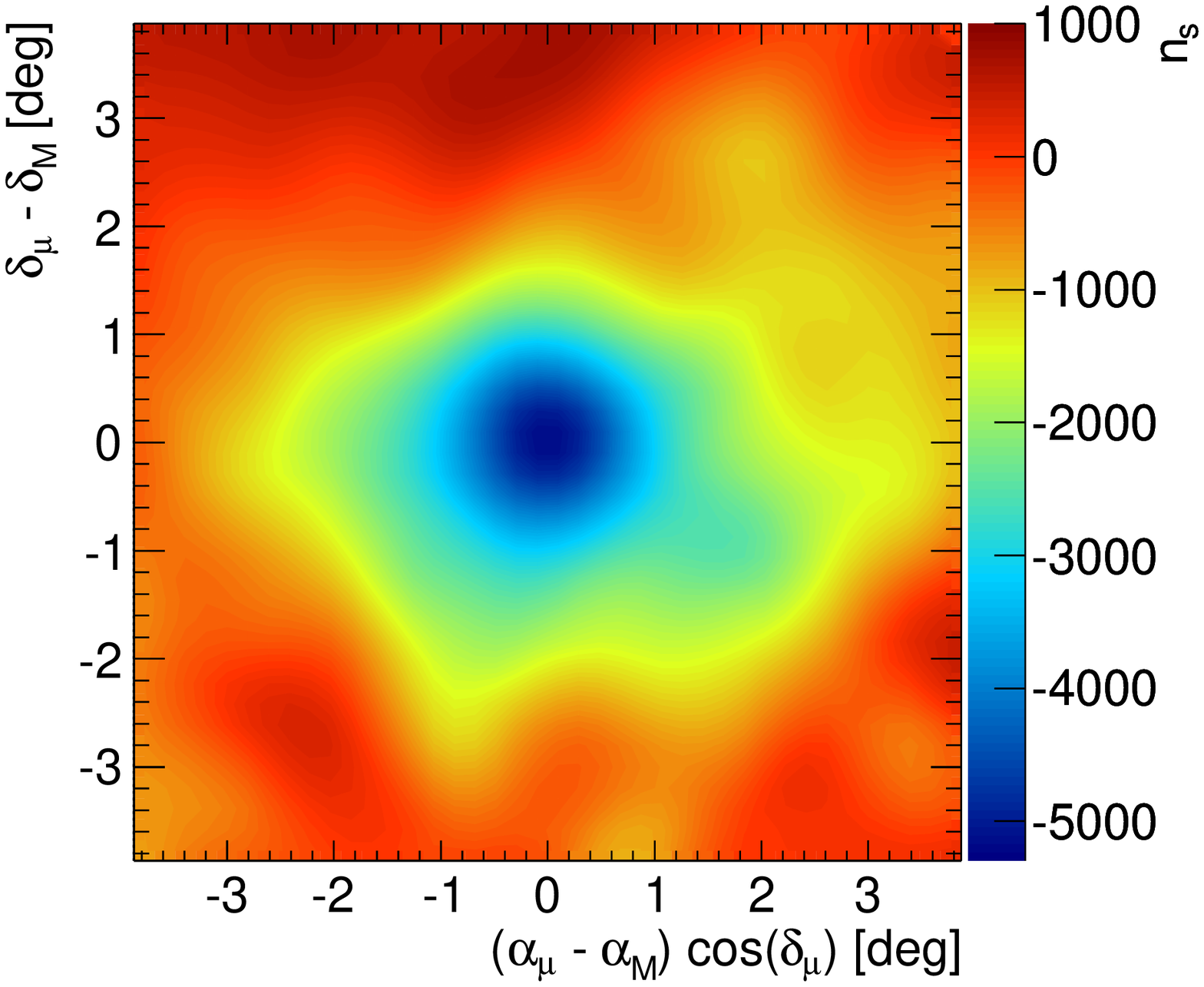} &
 \includegraphics[width=0.4\textwidth]{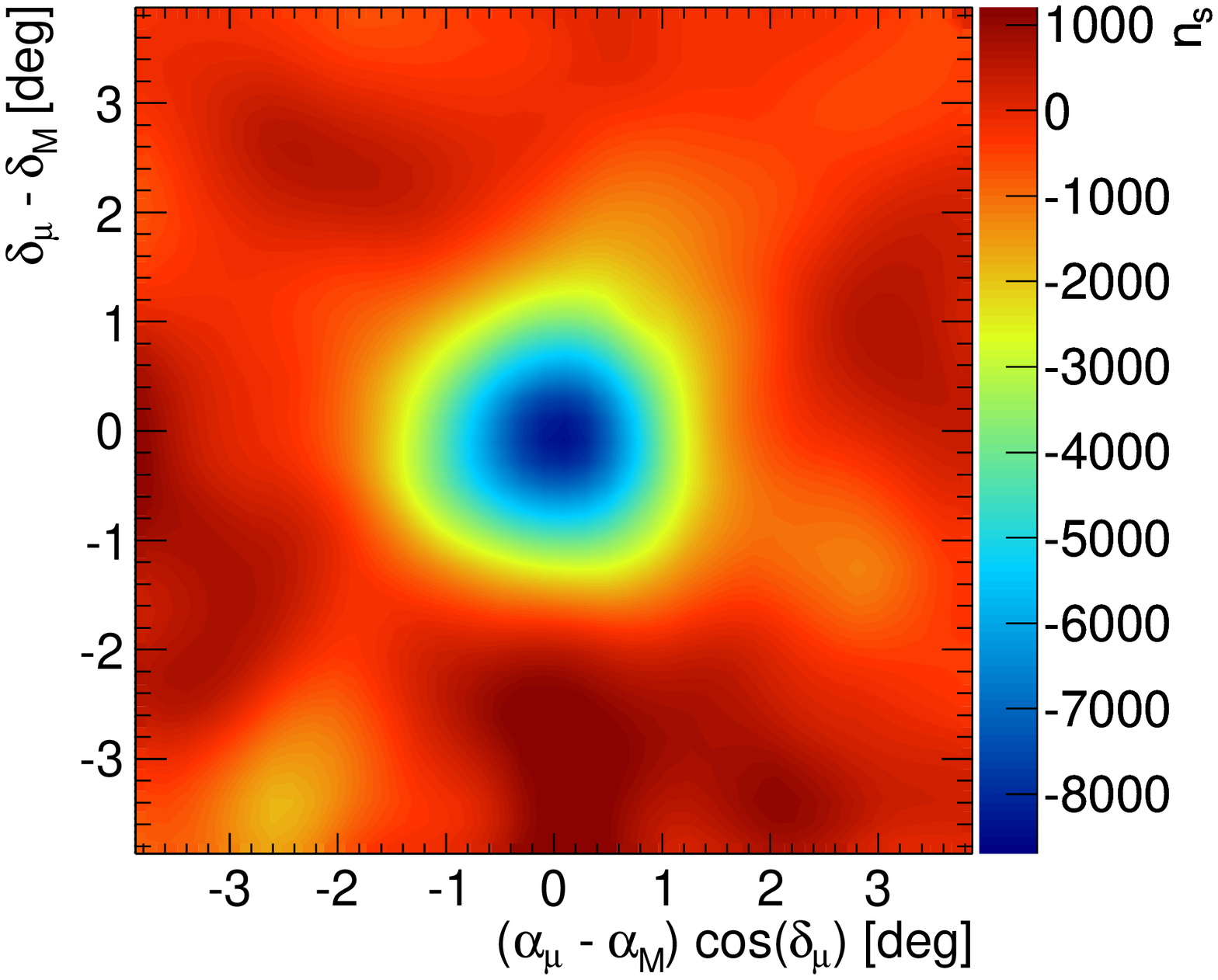} \\
 \end{array}$
 \end{center}
  \caption{Contour plot of the value of $n_s$ in the $(\Delta\alpha, \Delta\delta)$ coordinate system for
     \emph{on-source} regions of the IC40 \emph{(left)} and IC59 datasets \emph{(right)}.} 
    \label{fig:ic40_ic59_ns}
\end{figure*}

As described in the previous section, the maximum-likelihood values of $n_s$ were calculated on a
grid around the position of the Moon for both sets. The contour maps of the $n_s$ values obtained
for IC40 and IC59 are shown in Fig.~\ref{fig:ic40_ic59_ns}, where the shadowing effect of the Moon
is visible as a strong deficit in the central regions of the maps. The deepest deficit
observed with both detector configurations is in good agreement with the expected number of shadowed
events, listed in Table~\ref{tab_ub}. Using the RMS spread of the \emph{off-source} regions as a
$1\sigma$ estimator in the case of a null detection, we calculated the statistical significance of
the observation by taking the ratio of the largest deficit observed to the RMS spread, which is also
shown in Table~\ref{tab_ub}. The shadow of the Moon is observed in both the IC40 and IC59 datasets to high statistical 
significance ($> 10\sigma$).

\begin{figure*}[t]
  \begin{center}	
     $\begin{array}{cc} %
      \includegraphics[width=0.4\textwidth]{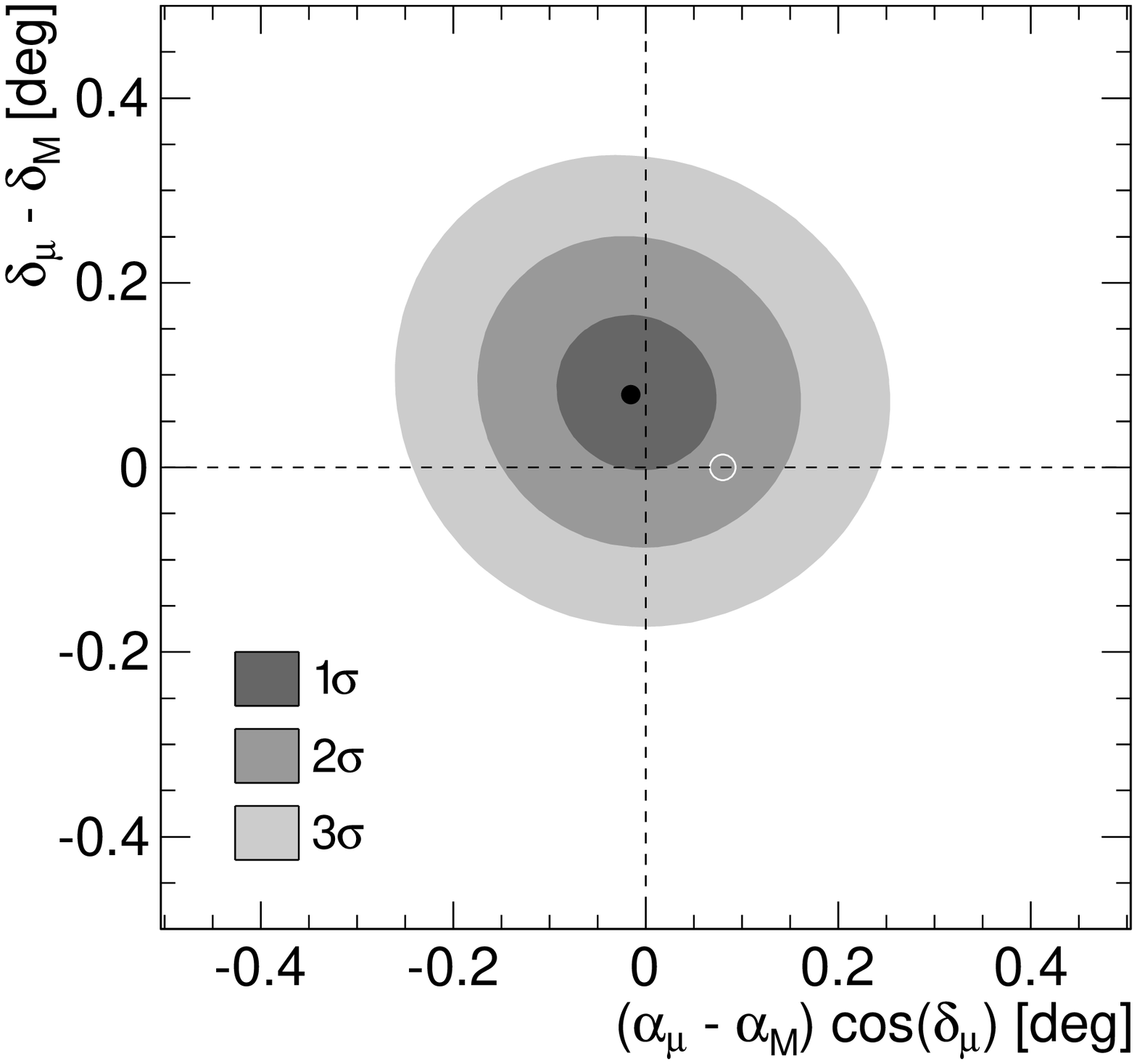} &
      \includegraphics[width=0.4\textwidth]{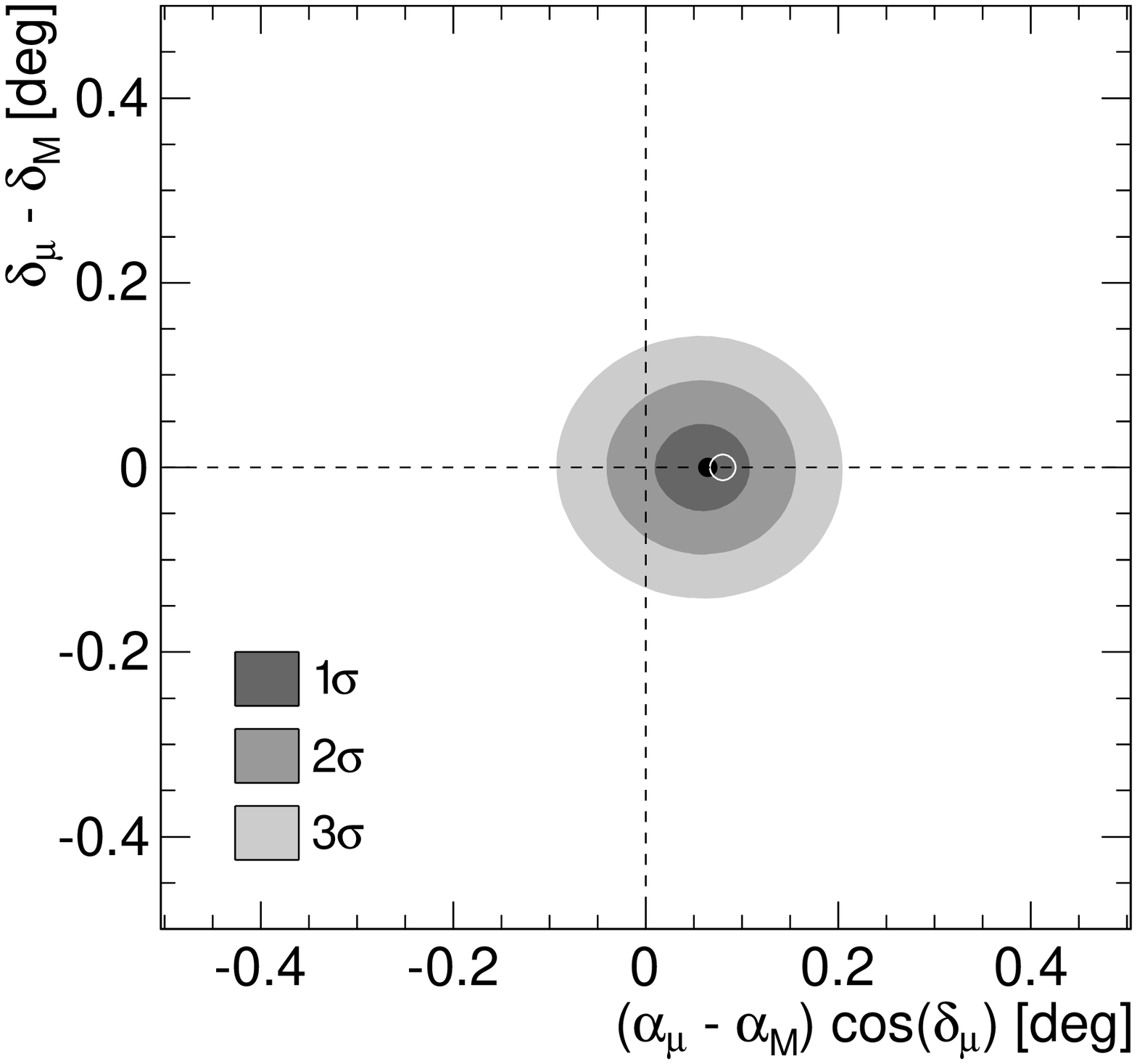}
     \end{array}$
    \caption{Contour plot for the position of the minimum of the Moon shadow in the IC40 
    (\emph{left}) and IC59 data (\emph{right}) in the $(\Delta\alpha, \Delta\delta)$ coordinate system. The reconstructed position for the Moon 
    shadow from the maximum likelihood analysis is shown as a black point, while the expected position 
    of the Moon shadow after accounting for magnetic deflection is shown as a white circle. }
    \label{fig:ic40_ic59} 
  \end{center}
\end{figure*}

In order to obtain a better estimate of the position of the minimum of the shadow, a finer grid with
a spacing of about $ 0.016^{\circ}$ was used in the central $\pm 0.4^{\circ} \times \pm 0.4^{\circ}$
region around the Moon. Using this grid, we obtain the positions indicated in Table~\ref{tab_ub} as
offsets in right ascension ($\Delta \alpha$) and declination ($\Delta \delta$) with respect to the
nominal position of the Moon in the sky. The shadow positions for both detector configurations are
shown in Fig.~\ref{fig:ic40_ic59} together with $1\sigma$, $2\sigma$, and $3\sigma$ contours. The
expected location of the minimum after accounting for geomagnetic deflection effects is also given
for comparison. In both detector configurations, the observed position of the minimum is consistent
with its expected location to within statistical fluctuations. These measurements imply that the absolute 
pointing accuracy of the detector during the IC40 and IC59 data-taking periods was better than 
about $0.2^{\circ}$.

\subsubsection{MPE analysis}\label{subsubsec:mperesults}

The unbinned analysis was also applied to IC59 events reconstructed using the MPE algorithm and its
corresponding angular error estimate described in Section~\ref{subsec:data}. In simulations, the MPE
fit performs better than the SPE reconstruction thanks to its more realistic description of the
arrival times of multiple photons at each DOM. However, at high energies the algorithm can be
confused by stochastic energy losses that occur along the muon track and are not described in the
likelihood function implemented in the MPE algorithm.  This usually results in an underestimation of
the angular uncertainty on the reconstructed direction of the track. In practice, this problem can
be solved by rescaling the average pull (the ratio between the real and estimated angular errors as
obtained from simulation studies) to unity.  The MPE version of the unbinned
analysis was used as a verification of this correction technique. 

Simulation studies indicate an average pull of 1.55 for the MPE reconstruction, versus 1.0 for
SPE. Without correcting for this underestimation of the angular error in the MPE fit, the Moon
shadow analysis resulted in a minimum value for $n_s$ of $3574 \pm 434$ shadowed events, differing
by more than 5 standard deviations from the expectation of $6373 \pm 80$. Redoing this analysis with
the angular error estimates rescaled by a factor of 1.55 resulted in fitted $n_s$ value compatible
with expectation, validating the pull correction method.

In neutrino analyses, where the range of muon energies is much larger than in the Moon analysis 
sample, the applied MPE pull correction is energy-dependent, instead of using only the average
value of the pull for the entire data set.

\vspace{8ex} 

\section{Conclusions} \label{sec:conclusions}

The shadow of the Moon in TeV cosmic rays has been detected to high significance ($>
6\sigma$) using data taken with the IC40 and IC59 configurations of the IceCube Neutrino
Observatory.
For both detector configurations, the observed positions of the shadow minimum show good agreement
with expectations given the statistical uncertainties. An important implication of this observation is that any systematic effects introduced by the detector geometry and the event reconstruction on the absolute pointing capabilities of IceCube are smaller than about $0.2^{\circ}$. 

The average angular resolution of both data samples was estimated by fitting a Gaussian function to the 
shadow profile. In both cases, the $1\sigma$ width of the Moon shadow was found to be about $0.7^{\circ}$,
which is in good agreement with the expected angular resolution based on simulation studies 
of down-going muons.

The total number of shadowed events estimated using the unbinned analysis is also consistent with 
expectations for IC40 and IC59. This provides an indirect validation of the angular uncertainty 
estimator obtained from the reconstruction algorithm.  This is especially relevant for the MPE analysis, 
where simulation studies indicate that the uncertainty estimator has to be rescaled in order to avoid 
underestimating the true angular error. Applying this correction factor to the data results in a number 
of shadowed events compatible with expectation.

Note that the value of the average angular resolution determined in this analysis is \emph{not}
a direct measurement of the point spread function to be used in searches for point sources of high-energy 
neutrinos. Rather, the agreement of this value with the value estimated from our simulations
should be seen as an experimental verification of our simulation and the methods used to estimate
the angular uncertainty of individual track reconstructions. This angular uncertainty depends on
several factors, in particular on the energy with which the muon traverses the detector. As the
energy distribution for neutrino analyses differs from that of the Moon shadow analysis, the average
angular resolution may be better or worse, but can reliably be estimated from our simulation.

\begin{acknowledgments}

We acknowledge the support from the following agencies:
U.S. National Science Foundation-Office of Polar Programs,
U.S. National Science Foundation-Physics Division,
University of Wisconsin Alumni Research Foundation,
the Grid Laboratory Of Wisconsin (GLOW) grid infrastructure at the University of Wisconsin - Madison, the Open Science Grid (OSG) grid infrastructure;
U.S. Department of Energy, and National Energy Research Scientific Computing Center,
the Louisiana Optical Network Initiative (LONI) grid computing resources;
Natural Sciences and Engineering Research Council of Canada,
WestGrid and Compute/Calcul Canada;
Swedish Research Council,
Swedish Polar Research Secretariat,
Swedish National Infrastructure for Computing (SNIC),
and Knut and Alice Wallenberg Foundation, Sweden;
German Ministry for Education and Research (BMBF),
Deutsche Forschungsgemeinschaft (DFG),
Helmholtz Alliance for Astroparticle Physics (HAP),
Research Department of Plasmas with Complex Interactions (Bochum), Germany;
Fund for Scientific Research (FNRS-FWO),
FWO Odysseus programme,
Flanders Institute to encourage scientific and technological research in industry (IWT),
Belgian Federal Science Policy Office (Belspo);
University of Oxford, United Kingdom;
Marsden Fund, New Zealand;
Australian Research Council;
Japan Society for Promotion of Science (JSPS);
the Swiss National Science Foundation (SNSF), Switzerland.

\end{acknowledgments}


%

\end{document}